\def\Tr{{\mathrm Tr}\ }
\def\j{{\mathbf j}}

\def\be{\begin{equation} }
\def\ee{\end{equation} }
\def\eq{\ =\ }

\def\JPCM{{\sl J. Phys. Condens Matter\ }}
\def\prb{{\sl Phys. Rev.\ }}
\def\PR{{\sl Phys. Rev.\ }}
\def\PRL{{\sl Phys. Rev. Lett.\ }}
\def\be{\begin{equation}}
\def\ee{\end{equation}}
\def\be{\begin{equation}}
\def\ee{\end{equation}}
\def\n{\noindent\ }
\documentclass[twocolumn,showpacs,superscriptaddress,preprintnumbers,amsmath,amssymb,prl]{revtex4}
\usepackage{graphicx}
\usepackage{dcolumn}
\usepackage{color}
\usepackage{bm}
 
  \def\ket{\vert \vert  \{ \emptyset \} \rangle}
  \def\ket2{\vert \vert \otimes \{ R \} \rangle}

  \def\ket{\vert \vert  \{ \emptyset \} \rangle}
  \def\ket2{\vert \vert \otimes \{ R \} \rangle}

\def\prl#1{ Phys. Rev. Lett. {\bf #1}}
\def\.#1{\mathaccent 95#1}
\def\^#1{\mathaccent 94 #1}
\def\~#1{\mathaccent "7E #1}

\def\eq{\enskip =\enskip}

  \def\ket{\vert \vert  \{ \emptyset \} \rangle}
  \def\ket2{\vert \vert \otimes \{ R \} \rangle}

\def\prl#1{ Phys. Rev. Lett. {\bf #1}}
\def\.#1{\mathaccent 95#1}
\def\^#1{\mathaccent 94 #1}
\def\~#1{\mathaccent "7E #1}

\def\eq{\enskip =\enskip}

  \def\ket{\vert \vert  \{ \emptyset \} \rangle}
  \def\ket2{\vert \vert \otimes \{ R \} \rangle}
  
\def\k{{\bf k}}

\def\etal{{\sl et.al.}\ }
\begin{document}

\title {Effect of short-range order on the electronic structure and optical properties of the CuZn alloy : an augmented space approach.}  
\author{Kartick Tarafder}
\affiliation{S.N. Bose National Center for Basic Sciences, JD Block, Sector III, Salt Lake City, Kolkata 700 098, India }
\author{Atisdipankar Chakrabarti \footnote{Permanent address : Ramakrishna Mission Vivekananda Centenary College, Rahara, West Bengal, India}}
\affiliation{S.N. Bose National Center for Basic Sciences, JD Block, Sector III, Salt Lake City, Kolkata 700 098, India }
\author{Kamal Krishna Saha}
\affiliation{Theory Department, Max-Planck-Institut f\"ur Mikrostrukturphysik,
Weinberg 2, D-06120 Halle (Saale), Germany}
\author{Abhijit Mookerjee}
\affiliation{S.N. Bose National Center for Basic Sciences, JD Block, Sector III, Salt Lake City, Kolkata 700 098, India }

\begin{abstract}
In this work we have combined the generalized augmented space method 
introduced by one of us with the recursion method of Haydock \etal (GASR),
 within the framework of the local density functional based linear muffin-tin
 orbitals basis (TB-LMTO). Using this we have studied the effect of short-range ordering and clustering on the density of states, optical conductivity and reflectivity of 50-50 CuZn alloys. Our results are in good agreement with alternative techniques. We argue that the TB-LMTO-GASR is a feasible, efficient and quantitatively accurate computational technique for the study of environmental effects in disordered binary alloys.
\end{abstract}
\date{\today}
\maketitle
\section{I. Introduction}
Binary alloys involving equal proportions of a noble metal Cu and a divalent metal Zn have a stable 
low temperature $\beta$-phase which sits immediately to the right of the pure face-centered cubic Cu phase in the alloy phase diagram \cite{jm,mas}. This phase,
 called $\beta$ brass, has a body-centered cubic structure. At high temperatures the alloy
forms a disordered body-centered cubic structure. At around 730K it orders into the
B2 structure with two atoms per unit cell. The alloy satisfies the Hume-Rother\'y
 rules \cite{hr} and has the same ratio of valence electrons to atoms. Jona and Marcus \cite{jm} have shown from a density functional theory (DFT) based approach that
within the local density approximation (LDA), it is the body-centered based B2 
which is the stable ground state. They also showed that if we include the gradient corrections (GGA) then we get a tetragonal ground state lower in energy by 0.1 mRy/atom.
This is in contradiction with the latest experimental data. The alloying of face-centered cubic Cu with an equal amount of Zn leads to a body-centered stable phase. Zn has only one more electron than Cu. This is an interesting phenomenon. CuZn alloys also have anomalously high elastic anisotropy. This makes the theoretical study of CuZn an interesting exercise for a proposed theoretical technique.

One of the  earliest first-principles density functional based study of the electronic properties of CuZn
was by Bansil \cite{bansil}. The authors  had  studied the complex bands of $\alpha$-phase of CuZn using the Korringa-Kohn-Rostocker (KKR) method coupled with the coherent potential approximation (CPA) to take care of disorder. They commented on the effects of charge transfer and lattice constants on the electronic structure. They found the  electronic distribution of this alloy to be of
 a split band kind with the centers of the Cu
and Zn $d$-bands well separated from each other. Their Zn $d$-bands showed hardly any dispersion and were shown only schematically in their figures. In a later work Rowlands \cite{row} generalized the CPA to a non-local version (NL-CPA) and studied the effects of short-range ordering in CuZn. Their technique was based on an idea of renormalization in reciprocal space suggested by Jarrell and Krishnamurthy \cite{jar}.  

The order-disorder transition in CuZn is a classical example of a true second order transition. Several very early investigations on this alloy have been reviewed by Nix and Shockley \cite{nix} and Guttman \cite{gutt}. These investigations, of course, were rather crude, since sophisticated approaches 
to deal with disordered alloys had really not been developed at that time.  However, it was recognized that a knowledge of the short-range order correlations above the critical temperature should be
of considerable interest. Early neutron scattering experiments were carried out on $\beta$-brass
by Walker and Keating \cite{wk}.  The Warren-Cowley short-range order parameter, defined by
$\alpha(R)=1-P_{AB}(R)/x$, where x was the concentration of A and $P_{AB}(R)$ was
 the probability of finding an A atom at a distance  of R from a B atom,  was directly obtained from the diffuse scattering cross-section :

\[ \frac{d\sigma}{d\Omega}\ =\ x(1-x)(b_A-b_B)^2 \sum_R \alpha(R) f(K)\exp(iK\cdot R)\]

$b_A,b_B$ were the scattering lengths of $A$ and $B$ atoms, and $f(K)=\exp(-C\vert K\vert^2)$ was the attenuation factor arising from thermal vibrations and static strains. The experimental data for
the short-range order parameter as a function of temperature are thus available to us.
A Ising-like model using pair interactions was studied by Walker and Chipman \cite{wc} and the
short-range order was theoretically obtained. However, the pair interactions were simply fitted
to the experimental values of the transition temperature $T_C$ and in that sense it was an
empirical theory. The experimental estimate of the nearest neighbour Warren-Cowley parameter
was found to be varying between -0.171 to -0.182 at around 750K. 

In a later work using the much more sophisticated {\sl locally self-consistent Green function} (LSGF)
approach based on the tight-binding linear muffin-tin orbital (TB-LMTO) technique Abrikosov \etal
 \cite{abri} studied CuZn alloys. The authors argued that earlier studies of the mixing enthalpies of CuZn using the standard coherent potential approximation approaches \cite{cpa1}-\cite{cpa5}
showed significant discrepancies with experiment.  The discrepancies were  assumed to partly 
arise from the neglect of charge transfer effects and partly because of short ranged ordering (SRO).
The main thrust of this technique, which was based on an earlier idea of a locally self-consistent multiple scattering (LSMS) by Wang \etal \cite{wang}, was to go beyond the CPA and 
include the effects of the immediate environment of an atom in the solid.  The LSMS gave an excellent
theoretical estimate of the ordering energy in CuZn : 3.37 mRy/atom as compared to the experimental
value of 3.5 mRy/atom.  The LSGF approach correctly predicted ordering tendency in CuZn on lowering 
temperature and combining with a Cluster Variation-Connolly Williams (CVM-CW) obtained a value of the nearest neighbour 
Warren-Cowley SRO parameter $\alpha$ = -0.15.  Subsequently Bruno \etal \cite{bruno} proposed a modification of the CPA including the local field effects 
and showed that charge transfer effects can be taken into account as accurately as  the O(N) methods just described. They applied their approach to the CuZn alloys.

One of the earlier works on the optical property of CuZn alloy was the determination of the temperature variation of optical reflectivity by Muldawer \cite{mul}. The author attempted to explain
the color of the disordered $\beta$-brass CuZn alloy via the internal photoelectric effect \cite{jk}.
Although, the experimental data also contained the contribution from plasma oscillations, the author claimed that the optical reflectivity helps to  explain the band picture of the alloys as a function of the inter-atomic spacing. In order to explain the optical properties, Amar \etal \cite{amar1}-\cite{amar3} studied the band structure of CuZn using the KKR method. However, they had used the virtual crystal approximation, replacing the random potential seen by the electrons by an averaged one. This is now known to be particularly inaccurate for split band alloys. 

The above discussion was necessary to bring into focus the following points~: in the study of alloys like CuZn it would be interesting to address the effects of charge transfer and short-range ordering. In this communication we shall address exactly these two points. We shall propose the use of the augmented space recursion (ASR) coupled with the tight-binding linear muffin-tin orbitals basis (TB-LMTO) \cite{asr}
to study the effects of short-range ordering on both the electronic structure and the optical properties of $\beta$-CuZn alloy at 50-50 composition.
We should like to stress here that the TB-LMTO-ASR addresses precisely these
effects with accuracy : the density functional self-consistent TB-LMTO takes care of the charge transfer, while the local environmental effects which are essential for the description of SRO are dealt with by the ASR. 
The TB-LMTO-ASR and its advantages  has been extensively discussed earlier in a review by Mookerjee \cite{tf} and in a series of articles  \cite{asr,Am,KG,asr2,Am3,Am4}. We would like to refer the interested readers to these for details. 

\section{II. Spectral functions, Complex bands and Density of States for 50-50 CuZn}

In this section we shall introduce the salient features of the ASR  which will be required by us in our
subsequent discussions. 

We shall start from a first principle TB-LMTO set of 
orbitals  \cite{Ander1,Ander2} in the most-localized
 representation. This is necessary, because the subsequent
recursion requires a sparse representation of the Hamiltonian. 
The TB-LMTO second order tight-binding Hamiltonian {\bf H}$^{(2)}$is described by a set of {
\sl potential parameters} :  {\bf C}$_R$, {\bf E}$_{\nu R}$ , ${\mathbf\Delta}_R$  and ${\mathbf o}_R$ which are characteristic of the atoms which sit on the
lattice sites labelled by $R$, and a 
 structure matrix  {\bf S}$_{RR'}$ which is characteristic of the lattice on
which the atoms sit. For a substitutionally disordered alloy, the structure
matrix is not random but the potential parameters are and can be described by a
set of random {\sl occupation variables} $\{n_R\}$. We may write~:

\[
{ C}_{RL}  =   C_L^A\  n_R + C_L^B\ \left( 1-n_R \right) 
\]

\noindent and similar expressions for the other potential parameters.
The random site-occupation variables $\{n_R\}$ take values 1 and 0 
 depending upon whether the muffin-tin labelled by $R$ is occupied by $A$ or
$B$-type of atom. The atom sitting at $\{R\}$ can either be of type $A \ (n_R=1)$
with probability $x$ or $B \ (n_R=0)$ with probability $y$.

In the absence of short-range order, the augmented space formalism associates with each random variable  $n_R$  an operator  ${\bf M}_R$
whose spectral density is its probability density.
\[
p(n_R) = -\frac{1}{\pi} \lim_{\delta\rightarrow 0} \ \mbox{Im} 
\langle \uparrow_R|\left((n_R+i\delta) {\bf I}-{\bf M}_R\right)^{-1}|\uparrow_R\rangle
\]
The operator {\bf M}$_R$ acts on the ``configuration space" of the variable $n_R$, $\mathbf{\Phi}_R$ spanned by
the configuration states $\vert \uparrow_R\rangle$ and $\vert \downarrow_R\rangle$. The augmented space theorem \cite{Am} states that a configuration average can be expressed as
a matrix element in the ``configuration space" of the disordered system : 

\be 
\ll A(\{ n_{R}\}) \gg \eq < \{\emptyset\}\vert \widetilde{{\mathbf A}}(\{{\mathbf M}_R\})\vert \{\emptyset\}>
\label{eq5}
\ee

\n where,

\[ \widetilde{\mathbf A}(\{{\widetilde{\bf M}_R}\}) \eq \int \ldots \int A(\{\lambda_{R}\})\ \prod d{\mathbf P}(\lambda_{R}).\]

\n {\bf P}($\lambda_{R}$) is the spectral density of the self-adjoint operator $\widetilde{{\bf M}}_{R}$, and the configuration state $\vert\{\emptyset\}\rangle$ is $\prod^\otimes_R \vert \uparrow_R\rangle$.
Applying (\ref{eq5}) to the Green function we get~:
\begin{equation}
\ll {\bf G}(\k,z)\gg \ = \ \langle \k\otimes\{\emptyset\} |{(z{\bf\widetilde I} 
- {\bf\widetilde{H}}^{(2)})}^{-1} |\k\otimes\{\emptyset\} \rangle.
\label{eq6}
\end{equation}

where {\bf G} and {\bf H}$^{(2)}$ are operators which are matrices in angular momentum space, and 
the augmented {\bf k-} space basis $|\k,L\otimes\{\emptyset\} \rangle$ has the form

\[ (1/\sqrt{N})\sum_R \mbox{exp}(-i\k\cdot R)|R, L\otimes \{\emptyset\}\rangle. \]

The augmented space Hamiltonian ${\bf\widetilde H}^{(2)}$ is constructed from the TB-LMTO
Hamiltonian ${\bf H}^{(2)}$ by replacing each random variable $n_R$ by  the operators $\widetilde{\bf M}_R$.
It is an operator in the augmented space 
$\Psi$ = ${\cal H} \otimes \prod^\otimes_R \mathbf{\Phi}_R$. The ASF maps 
a disordered Hamiltonian described in a Hilbert space ${\cal H}$ onto an ordered Hamiltonian 
in an enlarged space $\Psi$, where the space $\Psi$ is constructed as the outer product of the
space ${\cal H}$ and configuration space $\Phi$ of the random variables of the disordered 
Hamiltonian. The  configuration space $\Phi$ is of rank 2$^{N}$ if there are $N$ muffin-tin 
spheres in the system. Another way of looking at ${\bf\widetilde H}^{(2)}$ is to note
that it is the {\sl collection} of all possible Hamiltonians for all possible
configurations of the system.

This equation is now exactly in the form in which recursion method may be applied. At this point
we note that the above expression for the averaged $G_{LL}(\k,z)$ is {\sl exact}.

The recursion method addresses inversions of infinite matrices of the type associated
with the Green function \cite{vol35}. Once a sparse representation
of an operator in Hilbert space, ${\bf\widetilde H}^{(2)}$, is known in a countable basis, the recursion method
obtains an alternative basis in which the operator becomes tridiagonal. This basis and the 
representations of the operator in it are found recursively through a three-term recurrence relation.
The spectral function is then obtained from the 
 continued fraction~:

\begin{widetext}
\begin{equation}
\ll G_{LL}(\k,z) \gg \ = \ \frac{{\beta_{1L}^2}}
        {\displaystyle z-\alpha_{1L}(\k)-\frac{\beta^2_{2L}(\k)}
        {\displaystyle z-\alpha_{2L}(\k)-\frac{\beta^2_{3L}(\k)}
        {\displaystyle \frac{\ddots}
        {\displaystyle z-\alpha_{NL}(\k)-\mathbf \Gamma_L(\k,z)}}}}
\ = \ \frac{{\beta_{1L}^2}}{\displaystyle z-E_{L}(\k)-\Sigma_L(\k,z)}
\end{equation}
\end{widetext}

where $\mathbf \Gamma_L(\k,z)$ is the asymptotic part of the continued fraction.  
The approximation involved has to do with the termination of this continued fraction. The coefficients 
are calculated exactly up to a finite number of steps $\{\alpha_n,\beta_n\}$ for $n < N$ and the asymptotic
part of the continued fraction is obtained from the initial set of coefficients using the idea of Beer and
Pettifor terminator \cite{bp}. Haydock and coworkers \cite{kn:hay} have carried out extensive studies of 
the errors involved and precise estimates are available in the literature. Haydock \cite{kn:hay2} has shown
that if we carry out recursion exactly up to $N$ steps, the resulting continued fraction maintains the first 
$2N$ moments of the exact result.
\begin{figure*}
\centering
\includegraphics[width=2in,height=3in]{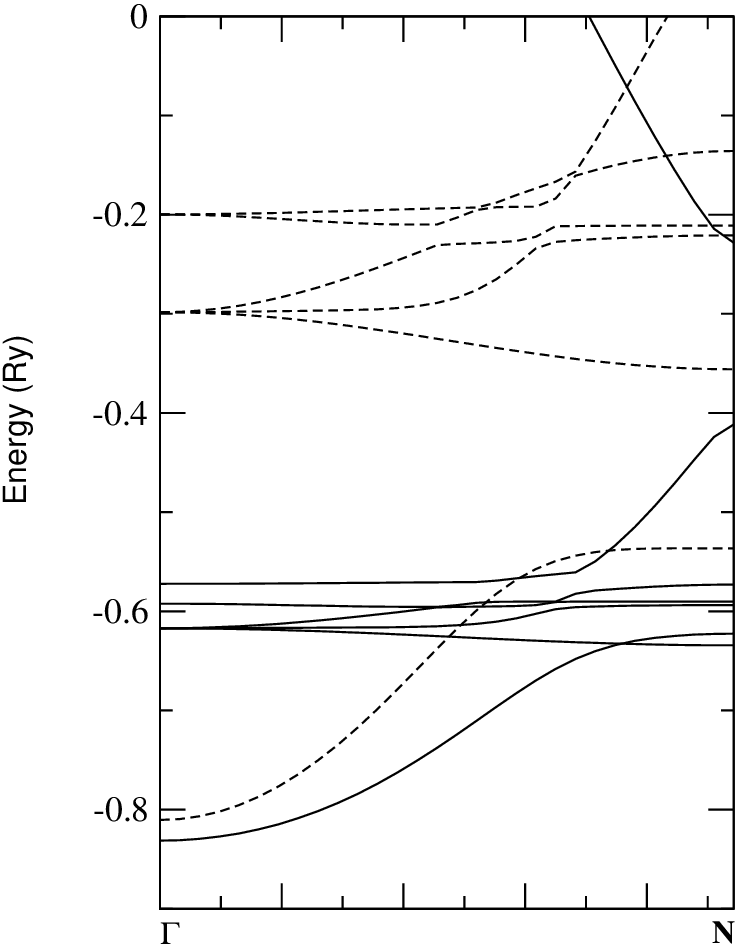} \hskip 1in
\includegraphics[width=2in,height=3in]{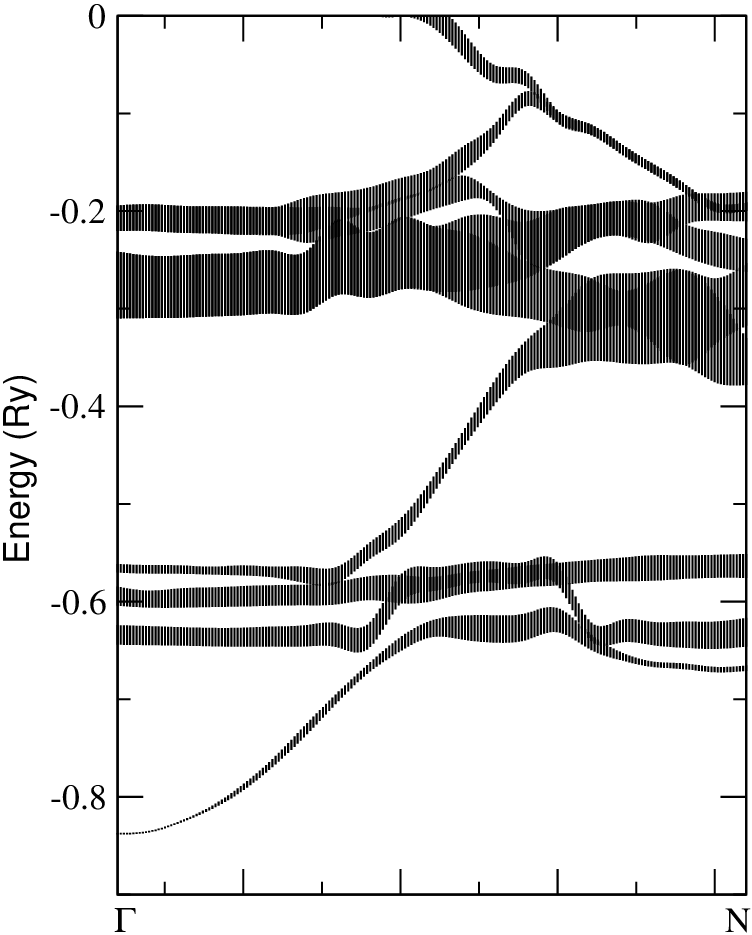}
\caption{(left) Bands for pure Cu and Zn in bcc lattices with the same
lattice parameter as the 50-50 CuZn alloy. The dashed line are for Cu and the full lines for Zn. (right) Complex bands for the
50-50 CuZn alloy.}
\label{fig1}
\end{figure*}

The self-energy $\Sigma_L(\k,z)$ arises because of scattering by the random potential fluctuations. 

The average spectral function $\ll A_\k(E) \gg$ is related to the averaged Green function in 
reciprocal space as :

\[ \ll A_\k(E) \gg \ = \sum_L \ll A_{\k L}(E) \gg ,\]
where
\[ \ll A_{\k L}(E) \gg \ = -\frac{1}{\pi} \lim_{\delta \rightarrow 0+} \{\mbox{Im} \ll G_{LL}(\k,E-i\delta)\gg\}. \]

\begin{figure}[b]
\vskip 0.3in
\centering
\includegraphics[width=2.5in,height=2.5in]{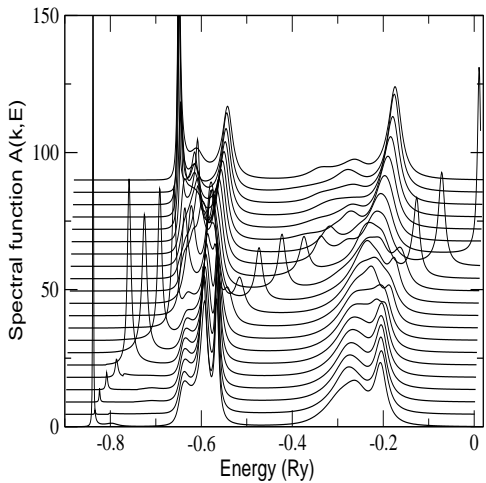}
\caption{Spectral functions for the CuZn alloy for {\bf k}-vectors along the $\Gamma$ to $N$ direction in the Brillouin zone.}
\label{fig2}
\end{figure}

To obtain the complex bands for the alloy we fix a value for $\k$ and solve for~:
\[ z-E_L(\k)-\Sigma_L(\k,E) = 0. \]
\begin{figure}
\centering
\includegraphics[width=4.5in,height=4in]{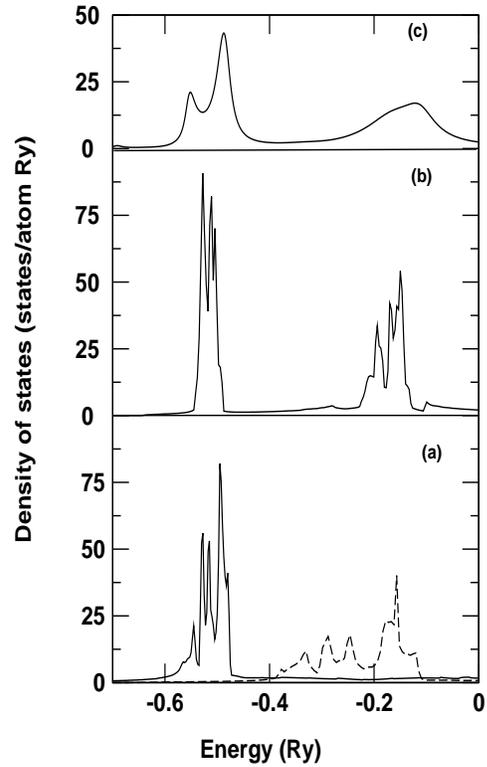}
\caption{(bottom) Density of states of pure Zn (solid lines) and Cu (dashed lines)in the same bcc lattice as the 50-50 CuZn alloy. (centre) Density of states
for ordered B2 50-50 CuZn alloy. (top) Density of states for the disordered
bcc 50-50 CuZn alloy. These results are comparable to the single-site CPA.} 
\label{fig3}
\end{figure}

The real part of the roots will give the position of the bands, while the imaginary part of
roots will be proportional to the disorder induced broadening. Since the alloy is random, the bands always have
finite lifetimes and are fuzzy.  

We have used this reciprocal space ASR to obtain the  complex bands and spectral functions for the CuZn alloy. This is shown in  Figs. (\ref{fig1}) - (\ref{fig2}).
It should be noted that we have carried out a fully LDA self-consistent calculation using the TB-LMTO-ASR developed by us \cite{adc} to obtain the potential parameters. It takes care of the charge transfer effects. For the  Madelung energy part of the alloy calculation, we have chosen the approach of Ruban and Skriver  \cite{skriver}. 

The two panels of Fig. (\ref{fig1}) compare the band structures of pure Cu and pure Zn metals in the same bcc lattice as the 50-50 alloy. We note that the $s$-like bands of Cu and Zn stretch from -0.8 Ry,while the $d$-like states of Zn and Cu, whose degeneracies are lifted by the cubic symmetry of the bcc lattice are
more localized and reside in the neighbourhood of -0.6 Ry and between -0.3 to -0.2 Ry respectively. The complex bands of the solid clearly reflect the same
band structure. However, the bands are slightly shifted and broadened because of the disorder scattering of Bloch states in the disordered alloy. The broadening due to disorder 
scattering are maximum for the Cu $d$-like bands, less for the Zn $d$-like bands and minimum for the lower $s$-like bands. This is because Cu and Zn atoms do not present much fluctuation in the potential for the $s$-like states.

\begin{figure}
\centering \includegraphics[width=1.5in,height=1.5in]{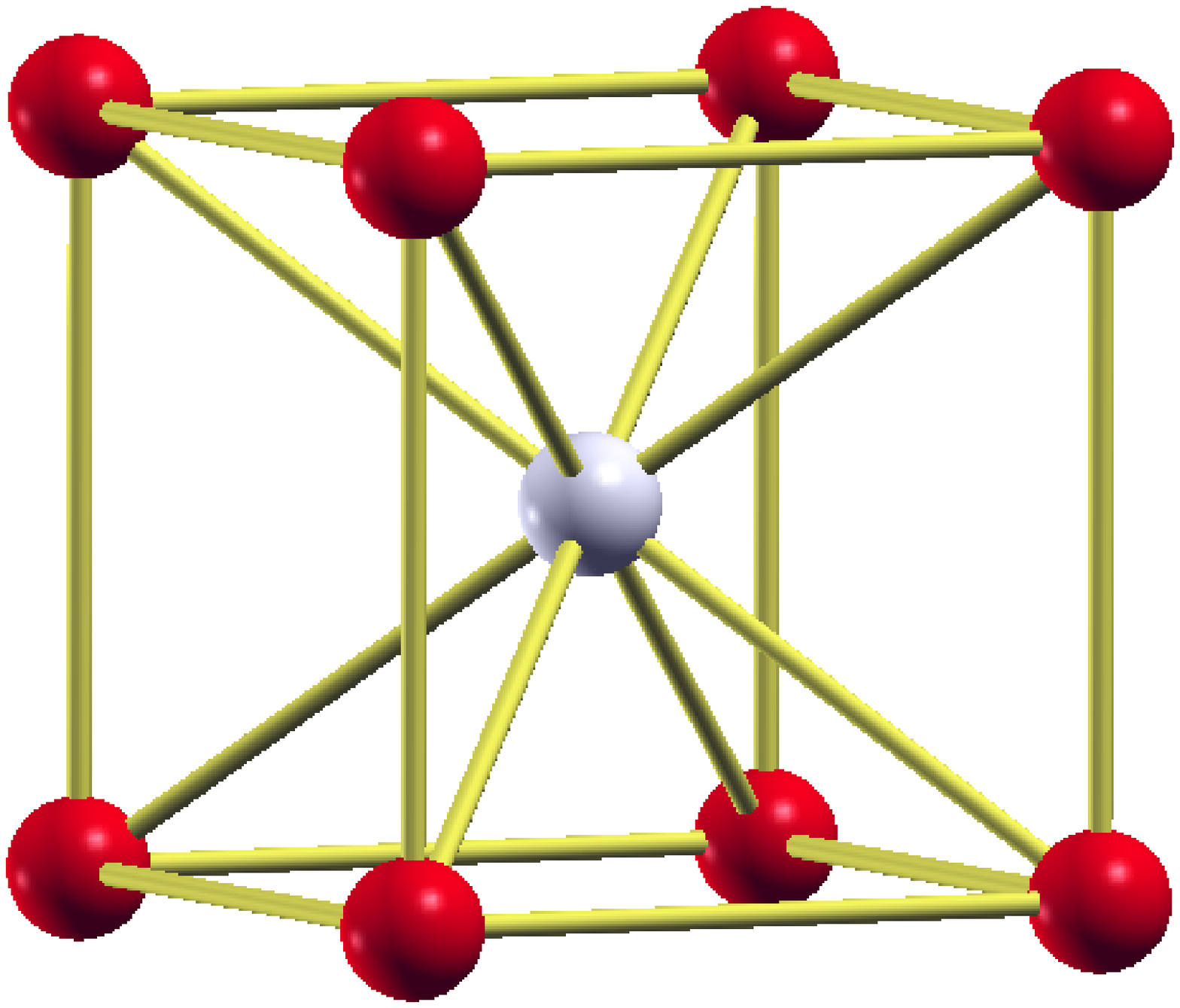}
\caption{(Color Online) (left) The light (blue) atom represents the site labelled 0, where the local density
of states will be calculated. The dark (red) atoms are its eight nearest neighbours on the bcc lattice.}
 \label{figa}
\end{figure} 

The same is reflected in the spectral functions, shown here also along the
$\Gamma$-$N$ direction in the Brillouin zone. Sharp peaks stretching from -0.8 Ry, groups of wider peaks around -0.6 Ry with less dispersion, characteristic of the more localized $d$-like states and groups of much wider peaks straddling
-0.3--0.2 Ry also with less dispersion.  The spectral functions play an important role in response functions related to photoemmission  and optical conductivity \cite{km1}. Our complex bands agree remarkably well with the Fig.3 of
Bansil and Ehrenreich \cite{bansil}. These authors of course did not show
the dispersion of the Zn $d$-bands, but as in their work, the Cu bands show
greater disorder induced broadening than the lower energy Zn $d$-bands.

 We may use the generalized tetrahedron method to pass from the reciprocal space spectral functions to the real space density of states \cite{kspace}. Alternatively, we may also carry out real-space ASR to obtain the density of states directly.

Fig. (\ref{fig3}) shows the densities of states for the pure Zn (solid lines) and Cu (dashed lines) in the same bcc lattice as the alloy and compares this with 
the ordered B2 and disordered bcc 50-50 CuZn alloy. We first note that in the ordered B2 alloy there is a considerable narrowing of the Zn  well as the Cu $d$-like bands. The feature around -0.35 Ry below the Fermi energy is suppressed in the ordered alloy. In the disordered alloy on the other hand, although disorder scattering introduces
life-time effects which washes out the sharp structures in the ordered systems,
the resemblance to the pure metals is evident. As seen in the complex bands, the life-time effects in the Cu $d$-like part is prominent. If we interpret the bottom most figure (a) as that due to completely segregated Cu-Zn and the middle one as the completely ordered one, then the disordered alloy lies between the two. In the next section, introducing short-range ordering effects on top of the fully disordered alloy, we shall study how to bridge between the two states.

\section{III. Short-Ranged Ordering in the alloys}
Attempts at developing generalizations of the coherent potential approximation (CPA) to include effects
of short-range order (SRO) have been many,  spread over the last several decades. 
The CPA being a single site mean-field approximation could not take into account SRO,
 since any description of SRO had to take into account correlations in, at least, a nearest neighbour cluster on the lattice. The early attempts to generalize the CPA to clusters 
 were beset with difficulties of violation of the analytic properties
of the approximated configuration averaged Green function.  Tsukada's \cite{mcpa} idea of introducing a 
super-cell of the size of the cluster immersed in an effective medium suffered from the problem of broken translational symmetry within the cluster even when the disorder was homogeneous. The
CCPA proposed by Kumar \etal \cite{ksm} based on the augmented space theorem also 
suffered from the same problem.  The embedded cluster
approximation of Gonis \etal \cite{gonis} immersed a cluster in a CPA medium which lacked the full self-consistency
with it.  The first translationally symmetric cluster approximations which preserved the analytic properties
of the approximate Green functions were all based on the augmented space theorem of Mookerjee \cite{Am}. They
included the travelling cluster approximation (TCA) of Kaplan and Gray \cite{KG} and Mills and Ratanavararaksa \cite{tca} and the CCPA proposed by Razee \etal \cite{razee}. The problem with these approaches was that they became intractable as the size of the cluster was increased much beyond two sites. Mookerjee and Prasad \cite{mp}
generalized the augmented space theorem to include correlated disorder. However, since they then went on to apply it in the CCPA approximation, they could not go beyond the two-site cluster and 
they applied the method to model systems alone. 
The breakthrough came with the augmented space recursion (ASR) proposed by Saha \etal \cite{sdm1}-\cite{sdm2}. The method was a departure from the mean-field approaches which always began by embedding a cluster
in an effective medium which was then obtained self-consistently. Here the Green function was expanded in a continued fraction whose asymptotic part was approximately estimated
 from its initial steps through an ingenious {\sl termination}
procedure \cite{vol35}. In this method the effect at a site of quite a large environment around it could be taken into account depending how far one went down the continued fraction before {\sl termination}.
The technique was made fully LDA-self-consistent within the tight-binding linear muffin-tin orbitals (TB-LMTO) 
approach \cite{atis} and several applications have been carried out to include short-range order in different alloy systems \cite{durga}. Recently Leath and co-workers have developed an itinerant CPA (ICPA) based on the
augmented space theorem \cite{glc}, which also maintains both analyticity and translational symmetry and takes into account effect of the nearest neighbour environment of a site in an alloy. The technique has been  successfully applied to the phonon
problem in alloys where there were large force constant disorders. 
The results of this method for NiPd and NiPt alloys match  well with the ASR applied
 to the same alloys \cite{alam} and there is now an  effort to apply the ICPA to
electronic problems based on both the TB-KKR and the TB-LMTO methods. A very different and rather striking
approach has been developed by Rowlands \etal \cite{row2} (the non-local CPA or NL-CPA) using the idea of {\sl coarse graining}
in reciprocal space originally proposed by Jarrell and Krishnamurthy \cite{jar}.  The NL-CPA with SRO  has been applied earlier by Rowlands \etal \cite{row} and 
is on the verge of being made 
 fully DFT self-consistent within the KKR. The authors report an unpublished report on it \cite{row4}.
In this communication we report a fully DFT self-consistent ASR based on the TB-LMTO with SRO incorporated.
We have applied it to the case of 50-50 CuZn alloys, so as to have a comparison with earlier attempts using different techniques.

\section{IV. The generalized augmented space theorem}

The generalized augmented space theorem has been described in detail by Mookerjee and Prasad \cite{mp}. Let us briefly introduce those essential ideas which are necessary to make this communication reasonably self-contained. 

For a substitutionally binary disordered alloy $A_xB_y$ on a lattice we can introduce a set of random {\sl occupation variables} $\{n_R\}$  associated with the lattice sites labelled by $R$, which take the values 0 or 1 depending upon whether the site $R$ is occupied by a A or a B type of atom. The Hamiltonian and hence the Green function are both functions of this set of random variables. 

To start with, let us assume that short-range order extends up to nearest neighbours only.
Let us take for an example the nearest neighbour cluster of nine atoms on a body centered cubic lattice (see Fig. \ref{figa}) centered on the site (light colored in the figure) 
labelled by $R_0$.
The occupation variables
 associated with its eight neighbours are correlated with $n_{R_0}$, but not with one another. Further none of the other occupation variables associated with more distant sites are
correlated with $n_{R_0}$.  

We may then write :

\begin{eqnarray*}
 P(n_{R_0},n_{R_1},\ldots n_{R_k}\ldots)  
  = \phantom{XXXXXXXXX}\\
\phantom{XXX}P(n_{R_0})\ \prod_{j=1}^8\ P(n_{R_j}\vert n_{R_0})\ \prod_{k>8}\ P(n_{R_k})
\end{eqnarray*}

The generalized augmented space theorem then associates with the random variables $\{n_{R_k}\}$
corresponding operators $\{\mathbf{M}_{R_k}\}$ in their configuration space. The construction of the representations of these operators has been discussed in detail in the paper by Mookerjee and Prasad \cite{mp}. Here we shall quote only the relevant results necessary to proceed further.

We shall characterize the SRO by a Warren-Cowley parameter $\alpha$. In terms of this the probability densities are given by :

\n For $k\ =\ 0$ and $k\ >\ 8$ 
\[
 P(n_{R_k}) =  x\ \delta(n_{R_k}-1)+y\ \delta(n_{R_k}), \quad x+y=1 
\]

\n For $1\ \leq j\ \leq 8$
\begin{eqnarray*}
P(n_{R_j}\vert n_{R_0}=1) &=& (x+\alpha y) \delta (n_{R_j}-1)+(1-\alpha)y \delta(n_{R_j}) \nonumber\\
P(n_{R_j}\vert n_{R_0}=0) &=& (1-\alpha)x\delta(n_{R_j}-1)+(y+\alpha x)\delta(n_{R_j}) 
\end{eqnarray*}

In the full augmented space, 
 the operators which replace the occupation variables are :

\begin{eqnarray}
\widetilde{\mathbf M}_{R_0}  =  M_{R_0}\otimes I\otimes I\otimes \ldots \quad\quad\phantom{XXXXXXXX} \nonumber \\
\widetilde{\mathbf M}_{R_j}  =\sum_{\lambda=0}^1 P^k_1\otimes M_{R_j}^\lambda \otimes I\otimes \ldots 
\quad \quad \mbox{\rm j=1,2,$\ldots$ 8} \nonumber \\
\widetilde{\mathbf M}_{k}   =  I\otimes I \otimes \ldots M_{R_k}\otimes I\otimes \ldots 
 \quad\quad \mbox{\rm k $>$ 8}\phantom{Xx} \nonumber \\
  P(n_{R_j}\vert n_{R_0}=\lambda) = \phantom{XXXXXXXXXXXXX}\nonumber\\
 -\frac{1}{\pi}\lim_{\delta\rightarrow 0} \mathrm{Im}
\langle\uparrow_{R_j}\vert \left((n_{R_j}+i\delta)-M_{R_j}^\lambda\right)^{-1}\vert 
\uparrow_{R_j}\rangle
\end{eqnarray}
 
\begin{figure*}
\centering
\vskip 0.8cm
\includegraphics[width=2in,height=3in]{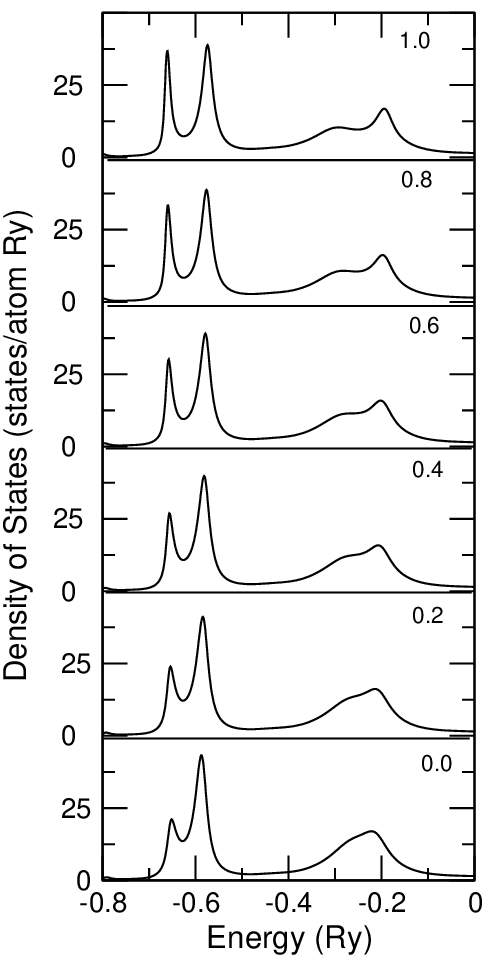}\hskip 2cm 
\includegraphics[width=1.8in,height=3in]{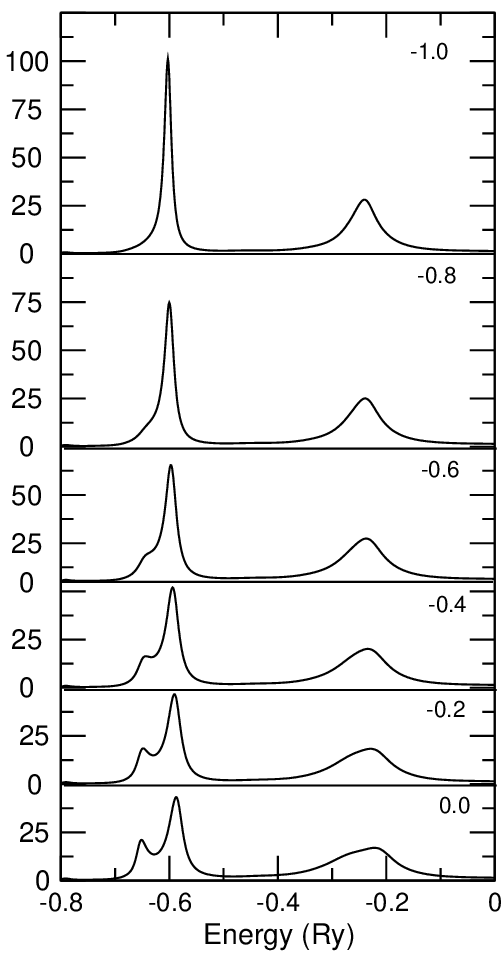}
\caption{Density of states for 50-50 CuZn with (left) increasing positive $\alpha$ which indicates
increasing clustering tendency and  (right) increasing negative $\alpha$ which indicates ordering tendency. The values of the SRO parameter $\alpha$ are shown on the upper right corner of each panel. Energies are shown with respect to the Fermi energy placed at the origin.}
\label{figsro}
\vskip 0.5in
\end{figure*}

We now follow the augmented space theorem and replace all the occupation variables $\{n_R\}$ by their corresponding operators. The configuration average is the specific matrix element between the {\sl reference} state $\vert \{\emptyset\}\rangle$ as discussed earlier. 
 We also note that the choice of the {\sl central} site labelled $R_0$ is immaterial. If we translate this site to any other and apply the lattice translation to
all the sites, the Hamiltonian in the full augmented space remains unchanged. This formulation of short ranged order also possesses lattice translational symmetry, provided the short-range order is homogeneous in space.

\section{V. Effect of SRO on the Density of States}

We have carried out the TB-LMTO-ASR  calculations on CuZn with a lattice constant of 2.85 \AA. The Cu and Zn potentials are 
obtained from  the LDA self-consistency loop. All reciprocal space integrals
are carried out  by using the generalized tetrahedron integration for disordered systems introduced by us earlier \cite{kspace}. 

 Let us discuss the effect of SRO, leading, on one hand to ordering ($\alpha < $ 0)
 and  on the other and to segregation ($\alpha >$ 0). We shall first look at Figs. \ref{fig1} and \ref{fig3}.  The complex band structure shown in Fig. \ref{fig1} shows
that the system is a {\sl split band} alloy. The positions of the $d$-bands of
Cu and Zn are well separated in energy. This implies that the ``electrons travel
more easily between Cu or between Zn sites than between unlike ones" \cite{row}. So when the alloy orders and unlike sites sit next to each other, the overlap
integral between the like sites decrease. This leads to a narrowing of the bands associated with Cu and Zn. A comparison between the bottom  and central panels of Fig. \ref{fig3} shows that the bands in the latter are much narrower than those of the former. This is the main effect of ordering setting in.  
 On the other hand, when the alloy is completely disordered, the bands gets widened by disorder scattering and the sharp structures in the density of states are smoothened. 

Fig. (\ref{figsro}) (left panel) shows the density of states with increasing positive $\alpha$
indicating increasing clustering tendency. Comparing with Fig. (\ref{fig3}) we note that as clustering tendency increases the density of states begins
to show the structures seen in the pure metals in both the split bands. For large positive $\alpha$ there is still residual long-ranged disorder. This causes smoothening of the bands with respect to the pure materials. For these large,
positive $\alpha$s, we notice the development of the structure around -0.35 Ry below the Fermi energy.

Fig. (\ref{figsro}) (right panel)  shows the density of states with increasing negative $\alpha$
indicating increasing ordering tendency. On the bcc lattice at 50-50 composition we expect this ordering to favour a B2 structure. With increasing ordering tendency, both the split bands narrow and lose structure. The feature around
-0.35 Ry disappears. This band narrowing and suppression of the feature around
-0.35 Ry is clearly seen in the ordered B2 alloy shown in Fig. (\ref{fig3}) (b).

Our analysis is closely similar to that of Rowlands \etal \cite{row}. 
Although there are differences in the way short-range order is introduced in
the ASR and NL-CPA, there is broad agreement between the two works on the effect of short-range ordering. In particular, the development of the shoulder around
-0.35 Ry below the Fermi energy with segregation and the narrowing of the split
bands on ordering are observed in both the approaches. In the ordering regime there are minor differences in the results of the two approaches. The relative heights of the two peaks in the split bands are much more pronounced and the broadening is larger in the ASR as compared with NL-CPA. 

\begin{figure}
\centering
\vskip 0.8cm
\includegraphics[width=2.5in,height=2.5in]{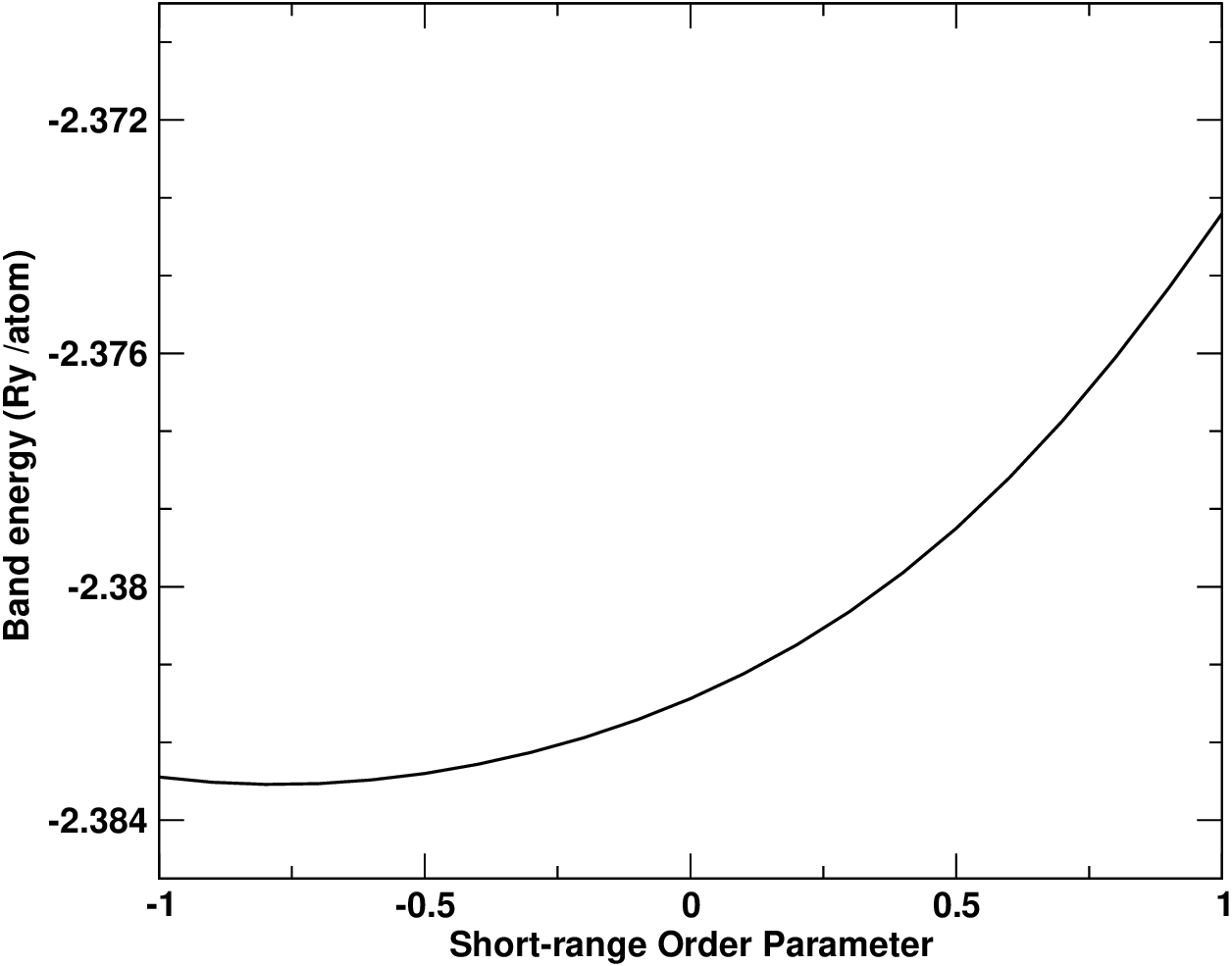}
\caption{The Band energy (from which the contribution of the core electrons
have been subtracted) as a function of the nearest-neighbour Warren-Cowley SRO parameter}
\label{stab}
\end{figure}

 Finally, in Fig. \ref{stab} we show the band energy as a function of the
nearest neighbour Warren-Cowley parameter. The minimum occurs at the ordering end, as expected. Experimentally the alloy does show a tendency to order at lower temperatures.

%%%%%%%%%%%%%%%%%%%%%%%%%%%%%%%%%%%%%%%%%%%%%%%%%%%
\section{VI. Optical properties of CuZn alloys}
In an earlier work \cite{tm} we had developed a methodology for the calculation of configuration averaged optical conductivity of a disordered alloy based on the augmented space formalism. Here we shall present the main features of the method necessary to understand the calculation of optical response functions in the 50-50 CuZn alloy. 

In linear response theory, at zero temperature, the generalized susceptibility of a disordered alloy is given by the Kubo formula : 

\[ \chi^{\mu\nu}(t-t') = (i/\hbar)\ \Theta(t-t')\ \langle[\j^\mu(t) \j^\nu(t')]\rangle \]

\n {\bf j}$^\mu$ is the current operator and $\Theta$ is the Heaviside step function.
If the underlying lattice has cubic symmetry, $\chi^{\mu\nu}$ = $\chi\ \delta_{\mu\nu}$.
The fluctuation-dissipation theorem then relates the imaginary part of the Laplace
transform of the generalized  susceptibility to the Laplace transform of a correlation
function :

\[
\chi^{\prime\prime}(\omega) = (1/2\hbar) \left(1-\exp\{-\beta\hbar\omega\}\right)S(\omega) \]
\n where,

\begin{equation}
 S(\omega) = \int_0^\infty\ dt \ \exp\{i(\omega + i\delta)t\}\ \Tr \left(\rule{0mm}{4mm} \j^\mu(t)\ \j^\mu(0)\ \right)
\end{equation}

\begin{figure*}
\centering
\vskip 1.5cm
\includegraphics[width=2.in,height=3in]{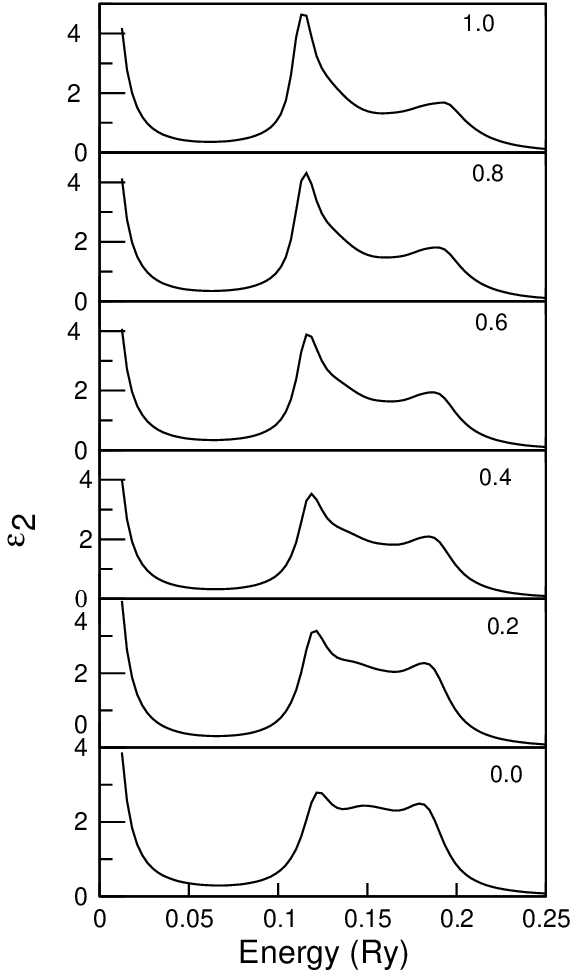}
\hskip 1cm
\includegraphics[width=2.in,height=3in]{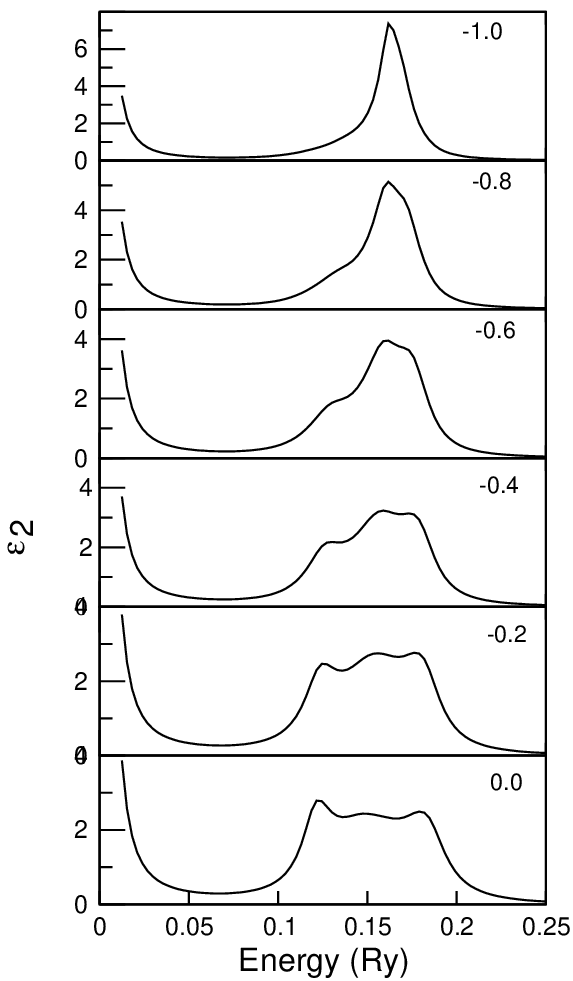}
\caption{(Left panels)Imaginary part of the dielectric function as a function of energy and its variation with the nearest neighbour Warren-Cowley SRO parameter,$\alpha\ >\ $ 0 increasing,  indicating
segregating tendency. (Right panels)
The same with $\alpha\ <\ $ 0 increasing, indicating ordering tendency. The value of $\alpha$ is given on the top right-hand corner of each panel.} 
\label{epsP}
\end{figure*}

Since the response function is independent of the direction label $\mu$ for cubic symmetry, in the
following we shall drop this symbol. In case of other symmetries we have to generalize our results
for different directions.
Our goal will be, given a quantum ``Hamiltonian" {\bf H}$^{(2)}$ to obtain the
correlation function,

\[
S(t) = \langle \phi\vert \mathbf{j}(t)\mathbf{j}(0)\vert\phi\rangle
\]

We shall now determine the correlation directly via the generalized recursion method as described by Viswanath and M\"uller\cite{gb}. 

For a disordered binary
alloy,  $S(t)= S[\bar{\mathbf H}(\{n_R\})]$. The augmented space theorem then
states that :

\begin{eqnarray*}
 \ll S(t)\gg   =  \ll \langle \phi\vert {\mathbf j}(t){\mathbf j}(0)\vert\phi\rangle \gg\phantom{XXXXXXXX}\\
  = \langle \phi\otimes \{\emptyset\}\vert \tilde{\mathbf j}(t)\tilde{\mathbf j}(0)\vert\phi\otimes\{\emptyset\}\rangle = S[\widetilde{\mathbf H}(\{\widetilde{\mathbf M}_R\})]
\end{eqnarray*} 

\n where the augmented space Hamiltonian and the current operators are constructed by replacing every random variable $n_R$ by the corresponding operator $\widetilde{\mathbf M}_R$.
 The recursion may now be carried out  step by step in the full augmented space and 
the configuration averaged structure function, which is the Laplace transform of the averaged correlation function can then be obtained as a continued fraction  :

\begin{equation}
\ll S(\omega)\gg \ =\ \lim_{\delta\rightarrow 0}\ 2\ \Re e\ \tilde{d}_0(\omega+i\delta) 
\label{eq6a}\end{equation}

\n where,
\begin{equation}
\tilde{d}_0(z) \ =\ \frac{i}{\displaystyle z-\tilde{\alpha}_0-\frac{\tilde{\beta}^2_1}{
\displaystyle{z-\tilde{\alpha}_1- \frac{\tilde{\beta}^2_2}{\displaystyle z-\tilde{\alpha}_2 - \ldots}}}}
\label{opt3}
\end{equation}

The imaginary part of the dielectric function is related to this correlation
function through :

\[ \epsilon_2(\omega)\ =\ \frac{\ll S(\omega)\gg}{\omega)} \]

The equations (\ref{eq6a})-(\ref{opt3}) will form the basis of our calculation of the configuration averaged correlation function.

\begin{figure*}[t]
\centering
\includegraphics[width=5.in,height=3.5in]{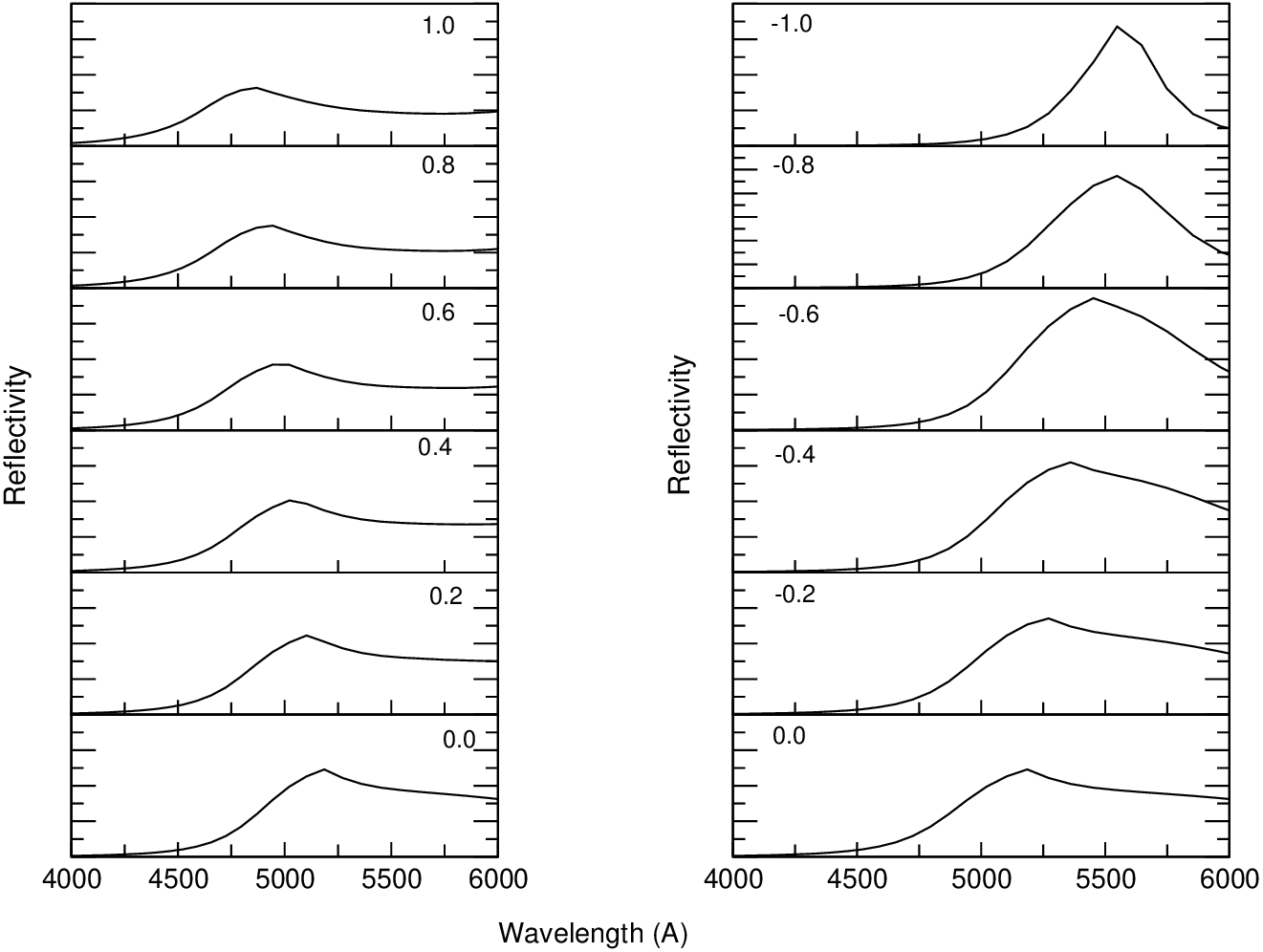}
\caption{(Left panels) Reflectivity as a function of the wavelength and its variation with the nearest neighbour Warren-Cowley parameter $\alpha\ >\ $ 0 increasing indicating segregating tendency. (Right panels)
The same with $\alpha\ <\ $ 0 indicating ordering tendency. The values of $\alpha$ are given on the top corners of each panel.} 
\label{ref}
\end{figure*}

The real part of the dielectric function $\epsilon_1(\omega)$ is related to the imaginary part $\epsilon_2(\omega)$ by a Kramer's Kr\"onig relationship. All optical response functions may now be derived from these. If we assume the orientation of the crystal surface to be parallel to the optic axis, the reflectivity $R(\omega)$ follows directly from Fresnel's formula :

\[ R(\omega)\ =\ \left | \frac{\sqrt{\epsilon(\omega)}-1}{\sqrt{\epsilon(\omega)}+1}\right |^2 \]

 Fig. \ref{epsP} shows the imaginary part of the dielectric function varying with frequency. If we examine the density of states for both the ordered and
disordered alloys we note that it is only around or just above $\hbar\omega \simeq$ 0.1 Ry that the transitions from the $d$-bands of Cu begin to contribute
to $\epsilon_2(\omega)$. Below this energy, the behaviour is Drude like. This is clearly seen in the panels of the figure. As $\alpha$ increases and the alloy
tends to segregate, since the weight of the structure in the density of states 
nearer to the Fermi-energy, increases, this contribution leads to the increasing
weight of the structure near 0.1 Ry. 

Fig. \ref{epsP} shows the variation of $\epsilon_2(\omega)$ with the nearest neighbour Warren-Cowley SRO parameter. The narrowing of the band with ordering and the increase of weightage to the structure away from the Fermi energy is clearly reflected in the dielectric function. It is also obvious from the figures that
the imaginary part of the dielectric function is more sensitive to the variation
of short-range order than the density of states.

Color changes in CuZn brasses have been observed both with changing composition,
hence different phase structures, and on heating \cite{mul}. In particular as these alloys are coloured and the changes can be observed visually. A study of
the changes in  optical reflectivity due to ordering or segregation  would be interesting. The {\sl colour} of these alloys should be due to the same physical mechanism of internal photoelectric excitations, as proposed for coloured metals like Cu and Au. Of course, we should take care of plasma effects and other
phenomena that together and in a combined way give rise to {\sl colour}. However, it would be, as an initial study, examine the effect of SRO on the optical 
reflectivity of the alloy.  This is sown in Fig. \ref{ref}.

The first thing we note that for the completely disordered alloy the reflectivity has a maximum around $\lambda\ =\ $ 5200\AA\ which is not very different from
that of Cu at low temperatures and is in the yellow-green region. Below this the reflectivity sharply drops (the reflectivity edge) and light of lower wavelengths is not reflected. With increasing segregation tendency this maximum drops almost linearly with $\alpha$ to around 4950\AA\ which is towards the green range. On the other hand with increasing ordering tendency the maximum moves out towards
5600\AA\ ,that is, moves towards the yellow from the yellow-green wavelengths. We would expect then a shift from a yellow-green colour in the disordered to a more
yellow region in the ordered material. We should note that these calculations are all done at 0K. In an actual experiment the disorder-order transition takes
place with varying temperature and the temperature effects on the electronic
structure must be taken into account, as well as effect of plasma oscillations.
We shall leave this for a later communication.

\section{VII. CONCLUSION AND COMMENTS}

The work presented here is a part of our continuing development of methods for the study of electronic structure of disordered alloys based on the augmented space method introduced by one of us. We have argued that our generalization of
the augmented space technique to include correlated disorder and its combination with the recursion method of Haydock \etal \cite{vol35}, yields configuration averaged Green functions which are lattice translationally symmetric and have the necessary herglotz analytical properties of the exact ones. 
We have applied this technique to the case of the split-band alloy 50-50 CuZn. Experimental evidences of short-range ordering in this alloy exist and hence our interest in its study. We have looked at the whole range of short-range ordering from clustering to homogeneous disorder to ordering and have studied its effect on the density of states, optical conductivity and reflectivity of the alloy. 
Our results are in broad agreement with an alternative approach via the non-local-CPA.  

We had set out to demonstrate that the generalized augmented space recursion, in
combination with the LDA based TB-LMTO, is an efficient computational technique which can go beyond the single-site mean-field approximation and take into account local environmental effects like short-range ordering and clustering, at the same time maintaining analytic properties essential for physical interpretation of its results. This is the main conclusion of this work.

\section*{Acknowledgment}
One of us (KT) would like to acknowledge the CSIR for financial assistance.

\end{document}